\documentclass[12pt,paper,notoc]{JHEP3}

\makeatletter
\usepackage{amssymb,epsfig}

\newcommand{\deriv}{\stackrel{\leftrightarrow}{D}}
\newcommand{\derleft}{\stackrel{\leftarrow}{D}}
\newcommand{\derright}{\stackrel{\rightarrow}{D}}

\def\bea{\begin{eqnarray}}
\def\eea{\end{eqnarray}}

\def\pl{{Phys. Lett.}~}

\def\lsim{ {\ \lower-1.2pt\vbox{\hbox{\rlap{$<$}\lower5pt\vbox{\hbox{$\sim$}
}}}\ } }
\def\gsim{ {\ \lower-1.2pt\vbox{\hbox{\rlap{$>$}\lower5pt\vbox{\hbox{$\sim$}
}}}\ } }

\title{Light-cone distribution amplitudes for the light $1^1P_1$
    mesons}

\author{Kwei-Chou Yang\\ Department of Physics, Chung Yuan Christian
University, Chung-Li, Taiwan 320, Republic of China}

\abstract{We present a study of light-cone distribution amplitudes
of the light $1^1P_1$ mesons. The first few Gegenbauer moments of
leading twist light-cone distribution amplitudes are calculated by
using the QCD sum rule technique.}

\preprint{hep-ph/0509337}

\begin{document}
\section{Introduction}
In the QCD description of various exclusive processes, it is
necessary to know the hadronic wave functions in terms of
light-cone distribution amplitudes (LCDAs). The role of LCDAs is
analogous to that of parton distributions in inclusive processes.
The conformal properties for multiplicative renormalizability of a
nonlocal operator, from which the defined wave functions are
written as a sum of LCDAs with specific conformal spins in the
asymptotic limit, have been systematically
studied~\cite{BF2,Braun:2003rp}. For each conformal spin, the
dependence of the distribution amplitudes on the transverse
coordinates is governed by the renormalization group equation and
the dependence on the longitudinal coordinates is involved in
"spherical harmonics" of the $SL(2,\mathbb{R})$ group. The
conformal invariance of QCD guarantees that for leading twist
LCDAs there is no mixing among Jacobi polynomials of different
spins to leading logarithmic accuracy, and, moreover, the
anomalous dimensions are ordered with conformal spin.

In the present work, we devote to the study of leading twist LCDAs
of $1^1P_1$ mesons: $b_1(1235), h_1(1170), h_1(1380)$, and
$K_{1B}$. We give a detailed calculations for Gegenbauer moments
of leading twist LCDAs. $h_1(1380)$ is a $1^1P_1$
meson\footnote{$h_1(1380)$ with $I^G(J^{PC})=?^-(1^{+-})$ was
denoted as $H'$ in old classification. Its isospin may be $0$, but
not confirmed yet.} and its properties are not experimentally
well-established~\cite{PDG}. The quark content of $h_1(1380)$ was
suggested as $\bar s s$ in the QCD sum rule
calculation~\cite{GRVW}. $K_{1B}$ are the $1^1P_1$ isodoublet
strange states. $K_{1B}$ and $K_{1A}$ (a $1^3P_1$ state) are the
mixtures of the real physical states $K_1(1270)$ and $K_1(1400)$,
where the mixing angle may be close to $45^\circ$~\cite{PDG}.

In the quark model, the $1^1P_1$ meson is represented as a
constituent quark-antiquark pair with total spin $S=0$ and angular
momentum $L=1$. Nevertheless, a real hadron in QCD language should
be described in terms of a set of Fock states for which each state
has the same quantum number as the hadron, and the leading twist
LCDAs are thus interpreted as amplitudes of finding the meson in
states with a minimum number of partons. Due to the G-parity, the
leading twist LCDA $\Phi_\parallel^A$ of a $1^1P_1$ meson defined
by the nonlocal axial-vector current is antisymmetric under the
exchange of $quark$ and $anti$-$quark$ momentum fractions in the
SU(3) limit, whereas the leading twist LCDA $\Phi_\perp^A$ defined
by the nonlocal tensor current is symmetric. The large magnitude
of the first Gegenbauer moment of $\Phi_\parallel^A$ could have a
large impact on the longitudinal fraction of
factorization-suppressed $B$ decays involving a $1^1P_1$ meson
evaluated in the QCD factorization framework~\cite{Yang:2005tv}.
Furthermore, $\Phi_\perp^A$ is relevant not only for exploring the
tensor-type new-physics effects in $B$ decays~\cite{Yang:2005tv}
but also for $B \to 1 ^1P_1 \ \gamma$ studies.

\section{Two-parton distribution amplitudes of $^1P_1$ axial
vector mesons}\label{sec:da}

We restrict ourselves to the two-parton LCDAs of the  $^1P_1$
axial vector mesons, denoted as $A$ here.\footnote{If the $^1P_1$
particle is made of $\bar q q$, then its charge conjugate $C$ is
$-1$, i.e., $J^{PC}=1^{+-}$.}. Throughout the present paper, we
define $z=y-x$ with $z^2=0$, and introduce the light-like vector
$p_\mu=P_\mu-m_A^2 z_\mu/(2 P\cdot z)$ with the meson's momentum
${P}^2=m_A^2$. The meson polarization vector
$\epsilon_\mu^{(\lambda)}$ is decomposed into {\it longitudinal}
and {\it transverse projections}, defined
as~\cite{Ball:1998sk,Ball:1998ff}
\begin{eqnarray}\label{app:polprojectiors}
 && \epsilon^{(\lambda)}_{\parallel\, \mu} \equiv
     \frac{\epsilon^{(\lambda)} \cdot z}{P\cdot z} \left(
      P_\mu-\frac{m_A^2}{P\cdot z} \,z_\mu\right), \qquad
 \epsilon^{(\lambda)}_{\perp\, \mu}
        = \epsilon^{(\lambda)}_\mu -\epsilon^{(\lambda)}_{\parallel\, \mu}\,,
\end{eqnarray}
respectively. In QCD description of hard processes involving axial
vector mesons, one encounters bilocal operators sandwiched between
the vacuum and the meson,
 \begin{eqnarray}\label{eq:general_op}
  \langle A(P,\lambda)|\bar q_1(y) \Gamma [y,x] q_2(x)|0\rangle ,
\end{eqnarray}
where $\Gamma$ is a generic notation for the Dirac matrix
structure and the path-ordered gauge factor is
\begin{equation}\label{eq:guagefactor}
[y,x]={\rm P}\exp\left[ig_s\!\!\int_0^1\!\! dt\,(x-y)_\mu
  A^\mu(t x+(1-t)y)\right].
\end{equation}
 This factor is equal to unity in the light-cone gauge which is
equivalent to  the fixed-point gauge (or called the Fock-Schwinger
gauge) $(x-y)^\mu A_\mu(x-y)=0$ as the quark-antiquark pair is at
the light-like separation. For simplicity, here and below we do
not show the gauge factor. For $b_1(1235)$ [$h_1(1170)$], the
operator in Eq.~(\ref{eq:general_op}) corresponds to $1/\sqrt{2}
(\bar u(y)\Gamma u(x)- [+] \bar d(y) d(x))$. In the present study,
we adopt the conventions: $D_\alpha=\partial_\alpha +ig_s
A^a_\alpha\lambda^a/2$, $\widetilde{G}_{\alpha\beta}=(1/2)
\epsilon_{\alpha\beta\mu\nu}G^{\mu\nu}$, $\epsilon^{0123}=-1$.

\subsection{Definitions}\label{subsec:da-def}
In general, the LCDAs are scheme- and scale-dependent. The
chiral-even LCDAs are given by
\begin{eqnarray}
  \langle A(P,\lambda)|\bar q_1(y) \gamma_\mu \gamma_5 q_2(x)|0\rangle
  && = if_A m_A \! \int_0^1
      du \,  e^{i (u \, p y +
    \bar u p x)}
   \left\{p_\mu
    \frac{\epsilon^{*(\lambda)} z}{p z} \, \Phi_\parallel(u)
         +\epsilon_{\perp\mu}^{*(\lambda)} \, g_\perp^{(v)}(u) \right. \nonumber\\
    &&\ \ \left. - \frac{1}{2}z_{\mu}
\frac{\epsilon^{*(\lambda)} z }{(p  z)^{2}} m_{A}^{2} g_{3}(u) \right\}, \label{evendef1} \\
  \langle A(P,\lambda)|\bar q_1(y) \gamma_\mu
  q_2(x)|0\rangle
  && = - i f_A m_A
\,\epsilon_{\mu\nu\rho\sigma} \,
      \epsilon^{*\nu}_{(\lambda)} p^{\rho} z^\sigma \, \int_0^1 du \,  e^{i (u \, p y +
    \bar u\, p  x)} \,
       \frac{g_\perp^{(a)}(u)}{4}, \label{evendef2}
\end{eqnarray}
with the matrix elements involving an odd number of $\gamma$
matrices and and $\bar u\equiv 1-u$. The chiral-odd LCDAs are
given by
\begin{eqnarray}
  &&\langle A(P,\lambda)|\bar q_1(y) \sigma_{\mu\nu}\gamma_5 q_2(x)
            |0\rangle
  =  f_A^{\perp} \,\int_0^1 du \, e^{i (u \, p y +
    \bar u\, p x)} \,
\Bigg\{(\epsilon^{*(\lambda)}_{\perp\mu} p_{\nu} -
  \epsilon_{\perp\nu}^{*(\lambda)}  p_{\mu})
  \Phi_\perp(u)\nonumber\\
&& \hspace*{+5cm}
  + \,\frac{m_A^2\,\epsilon^{*(\lambda)} z}{(p z)^2} \,
   (p_\mu z_\nu -
    p_\nu  z_\mu) \, h_\parallel^{(t)}(u)\nonumber\\
 && \hspace*{+5cm} + \frac{1}{2}
(\epsilon^{*(\lambda)}_{\perp \mu} z_\nu
-\epsilon^{*(\lambda)}_{\perp \nu} z_\mu) \frac{m_{A}^{2}}{p\cdot
z}
 h_{3}(u, \mu^{2})
\Bigg\},\label{odddef1}\\
&& \langle A(P,\lambda)|\bar q_1(y) \gamma_5 q_2(x) |0\rangle
  =  f_A^\perp
 m_{A}^2 \epsilon^{*(\lambda)} z\, \int_0^1 du \, e^{i (u \, p y +
    \bar u\, p x)}  \, \frac{h_\parallel^{(s)}(u)}{2},\label{odddef2}
\end{eqnarray}
with the matrix elements having an even number of $\gamma$
matrices. Here the LCDAs $\Phi_\parallel, \Phi_\perp$ are of
twist-2, $g_\perp^{(v)}, g_\perp^{(a)}, h_\perp^{(t)},
h_\parallel^{(s)}$ of twist-3, and $g_3, h_3$ of twist-4. In SU(3)
limit, due to G-parity, $\Phi_\parallel, g_\perp^{(v)}$,
$g_\perp^{(a)}$, and $g_3$ are antisymmetric under the replacement
$u\to 1-u$, whereas $\Phi_\perp, h_\parallel^{(t)}$,
$h_\parallel^{(s)}$, and $h_3$ are symmetric. Owing to the above
properties, we therefore normalize the distribution amplitudes to
be subject to
 \begin{equation}
 \int_0^1 du\Phi_\perp(u)=1,
 \end{equation}
and take $f_{A} =f_A^\perp(\mu=1$~GeV) in the study. Note that
\begin{equation}
 \int_0^1 du \Phi_\parallel(u)=\int_0^1 du g_\perp^{(a)}(u)
 =\int_0^1 dug_\perp^{(v)}(u)=\int_0^1 du g_3(u)=0
 \end{equation}
in SU(3) limit (see footnote~\ref{footnote:K1B} for further
discussions). We will not further discuss $g_3$ and $h_3$ below.

\subsection{Chiral-even light-cone distribution amplitudes}\label{subsec:da-even}

$\Phi_\parallel(u,\mu)$ can be expanded in a series of Gegenbauer
polynomials~\cite{BF2,Braun:2003rp}:
\begin{eqnarray}\label{eq:generaldaparallel}
\Phi_\parallel(u,\mu)=6u(1-u)\sum_{l=0}^\infty
a_l^{\parallel}(\mu) C^{3/2}_l(2u-1),
\end{eqnarray}
where $\mu$ is the normalization scale and the multiplicatively
renormalizable coefficients (or called Gegenbauer moments) are:
\begin{eqnarray}\label{eq:aparallel}
a_l^{\parallel}(\mu) = \frac{2(2l+3)}{3(l+1)(l+2)} \int_0^1 dx\,
C^{3/2}_l (2x-1) \Phi_\parallel(x,\mu).
\end{eqnarray}
In the limit of $m_{q_1}=m_{q_2}$, only terms with odd $l$ survive
due to G-parity invariance. In the expansion of
$\Phi_\parallel(u,\mu)$ in Eq.~(\ref{eq:generaldaparallel}), the
conformal invariance of the light-cone QCD exhibits that partial
waves with different conformal spin cannot mix under
renormalization to leading-order accuracy. As a consequence, the
Gegenbauer moments $a_l^{\parallel}$ renormalize multiplicatively:
  \begin{equation}
    a_l^{\parallel}(\mu) = a_l^{\parallel}(\mu_0)
  \left(\frac{\alpha_s(\mu_0)}{\alpha_s(\mu)}\right)^{-\gamma_{(l)}^\parallel/{b}},
  \label{eq:RGparallel}
   \end{equation}
where $b=(11 N_c -2n_f)/3$ and the one-loop anomalous dimensions
are \cite{GW}
  \begin{eqnarray}
  \gamma_{(l)}^\parallel  = C_F
  \left(1-\frac{2}{(l+1)(l+2)}+4 \sum_{j=2}^{l+1} \frac{1}{j}\right),
  \label{eq:1loopandim}
  \end{eqnarray}
with $C_F=(N_c^2-1)/(2N_c )$.

Applying the QCD equations of motion, discussed in detail in
Refs.~\cite{BB88,BF2,Ball:1998sk,Ball:1998ff}, one can obtain some
useful nonlocal operator identities such that the two-parton
disctribution amplitudes $g_\perp^{(a)}$ and $g_\perp^{(v)}$ can
be represented in terms of $\Phi_{\perp,\parallel}$ and two
three-parton distribution amplitudes. Neglecting three-parton
distribution amplitudes containing gluons and terms proportional
to light quark masses, $g_\perp^{(v)}$ and $g_\perp^{(a)}$ are
thus related to the twist-2 one by Wandzura-Wilczek--type
relations:
\begin{eqnarray}
 g_\perp^{(v)WW}(u)&\simeq&{1\over 2}\left[ \int_0^udv\,{1\over
\bar{v}}\Phi_\parallel(v) + \int_u^1dv\,{1\over
v}\Phi_\parallel(v) \right],
\label{eq:even-WW1}\\
g_\perp^{(a)WW}(u)&\simeq&2\bar{u} \int_0^udv\,{1\over
\bar{v}}\Phi_\parallel(v) + 2u\int_u^1dv\,{1\over
v}\Phi_\parallel(v). \label{eq:even-WW2}
 \end{eqnarray}

\subsection{Chiral-odd light-cone distribution amplitudes}\label{subsec:da-odd}

The leading twist LCDAs $\Phi_\perp^A(u,\mu)$ can be expanded
as~\cite{BF2,Braun:2003rp}
\begin{eqnarray}\label{eq:generaldaperp}
\Phi_\perp^A(u,\mu)=6u(1-u)\Bigg[1+ \sum_{l=1}^\infty
a_l^{\perp,A}(\mu) C^{3/2}_l(2u-1) \Bigg],
\end{eqnarray}
where the multiplicatively renormalizable Gegenbauer moments, in
analogy to Eq. (\ref{eq:aparallel}), read
\begin{eqnarray}\label{eq:aperp}
a_l^{\perp, A}(\mu) = \frac{2(2l+3)}{3(l+1)(l+2)} \int_0^1 dx\,
C^{3/2}_l (2x-1) \Phi_\perp^A(x,\mu)\,,
\end{eqnarray}
which satisfy
\begin{eqnarray}
    a_l^{\perp, A}(\mu) &=& a_l^{\perp, A}(\mu_0)
  \left(\frac{\alpha_s(\mu_0)}{\alpha_s(\mu)}\right)^{-(\gamma_{(l)}^\perp-
\frac{4}{3})/b}\,,
  \label{eq:RGperp}
   \end{eqnarray}
with the one-loop anomalous dimensions being~\cite{GW}
\begin{eqnarray}
  \gamma_{(l)}^\perp  = C_F
  \left(1+4 \sum_{j=2}^{l+1} \frac{1}{j}\right).
  \label{eq:gamma_perp}
  \end{eqnarray}
$a_l^{\perp, A}$ vanish in the SU(3) limit when $l$ are odd.

Using the equations of motion given in
Refs.~\cite{BB88,Ball:1998sk}, the two-parton distribution
amplitudes $h_\parallel^{(t)}$ and $h_\parallel^{(s)}$ can be
represented in terms of $\Phi_{\perp,\parallel}$ and a
three-parton distribution amplitude. The two-parton twist-3
distribution amplitudes are thus related to the twist-2 one
approximately by Wandzura-Wilczek--type relations
\begin{eqnarray}
h_\parallel^{(t)WW}(u) &=& \xi \left( \int_{0}^{u} dv
\frac{\Phi_{\perp}(v)}{\bar v} - \int_{u}^{1} dv
\frac{\Phi_{\perp}(v)}{v} \right),
\label{eq:3hww} \\
h_\parallel^{(s)WW}(u) &=& 2 \left( \bar u \int_{0}^{u} dv
\frac{\Phi_{\perp}(v)}{\bar v} + u \int_{u}^{1} dv
\frac{\Phi_{\perp}(v)}{v} \right). \label{eq:3eww}
\end{eqnarray}

\section{The tensor couplings and Gegenbauer
moments}\label{sec:coupling_moments}

\subsection{Input parameters}\label{subsec:inputs}
We calculate renormalization-group (RG) improved QCD sum
rules~\cite{SVZ} of the tensor couplings and Gegenbauer moments
for the $1^1P_1$ mesons. We employ the $SU(2)$ flavor symmetry,
i.e., do not distinguish between $b_1(1235)$ and $h_1(1170)$. We
therefore simply use $b_1$ to denote $b_1(1235)$ and $h_1(1170)$,
and $h_1$ to stand for the $h_1(1370)$ meson. In the numerical
analysis we take into account $\alpha_s(1~{\rm GeV})=0.517$,
corresponding to the world average
$\alpha_s(m_Z)=0.1213$~\cite{PDG}, and the following relevant
parameters at the scale $\mu=1$~GeV~\cite{Yang:1993bp}:
\begin{eqnarray}
\begin{array}{lcl}
  \langle \alpha_s G_{\mu\nu}^a G^{a\mu\nu} \rangle=0.474\ {\rm GeV}^4/(4\pi)\,, &  &   \\
  \langle \bar uu \rangle \cong \langle \bar dd \rangle =-(0.24\pm 0.005)^3~ {\rm
  GeV}^3 \,,
  &   & \langle \bar ss \rangle = 0.8 \langle \bar uu \rangle \,, \\
  (m_u+m_d)/2=5\ {\rm MeV}\,, &  & m_s=120\ {\rm MeV}\,,\\
  \langle g_s \bar u\sigma Gu \rangle \cong \langle g_s\bar
d\sigma Gd \rangle =-0.8\langle \bar uu \rangle, &  &\langle g_s
\bar s\sigma Gs \rangle = 0.8 \langle g_s\bar u\sigma Gu \rangle,
\end{array}\label{eq:parameters}
 \end{eqnarray}
with the corresponding anomalous dimensions of operators
satisfying~\cite{Yang:1993bp}:
 \begin{eqnarray}\label{app:anamolous}
 && m_{q, \mu}= m_{q , \mu_0}
  \left(\frac{\alpha_s(\mu_0)}{\alpha_s(\mu)}\right)^{-{4\over b}},
  \nonumber\\
 &&\langle \bar q q\rangle_\mu = \langle \bar q q\rangle_{\mu_0}
 \left(\frac{\alpha_s(\mu_0)}{\alpha_s(\mu)}\right)^{4\over b},\nonumber\\
 && \langle g_s \bar q\sigma\cdot G q\rangle_\mu =
 \langle g_s \bar q\sigma\cdot G q\rangle_{\mu_0}
 \left(\frac{\alpha_s(\mu_0)}{\alpha_s(\mu)}\right)^{-{2\over 3b}},\nonumber\\
 && \langle \alpha_s G^2\rangle_\mu = \langle \alpha_s
 G^2\rangle_{\mu_0}.
 \end{eqnarray}
We adopt the vacuum saturation approximation for describing the
four-quark condensates, i.e.,
\begin{eqnarray}
\langle 0|\bar q \Gamma_i \lambda^a q \bar q \Gamma_i \lambda^a
q|0\rangle =-\frac{1}{16N_c^2}{\rm Tr}(\Gamma_i\Gamma_i) {\rm
Tr}(\lambda^a \lambda^a) \langle \bar qq\rangle^2 \,,
\end{eqnarray}
and neglect the possible effects due to their anomalous
dimensions.

\subsection{The tensor couplings for $1^1P_1$ mesons}\label{subsec:coupling}
To determine the the tensor couplings of the $1^1P_1$ Mesons, $A$,
defined as
  \begin{equation}
  \langle 0 |\bar q_2 \sigma_{\mu\nu}q_1
  |A(P,\lambda)\rangle  = i f_A^\perp\,\epsilon_{\mu\nu\alpha\beta}
  \epsilon_{(\lambda)}^\alpha P^\beta\,,
  \label{eq:positivep1}
  \end{equation}
i.e.,
\begin{equation}
   \langle 0 |\bar q_2 \sigma^{\mu\nu}\gamma_5q_1 |A(P,\lambda)\rangle
 =-f_A^\perp (\epsilon_{(\lambda)}^{\mu} P^\nu -
\epsilon_{(\lambda)}^{\nu} P^\mu),
  \label{eq:positivep2}
\end{equation}
 we consider the correlation function of two tensor currents:
  \begin{eqnarray}
   \Pi_{\mu\nu\alpha\beta} &=&
  i\!\int\!\! d^4x\, e^{iqx} \,\langle 0|
  T[\bar q_1(x)\sigma_{\mu\nu} q_2(x)\, \, \bar q_2(0)
  \sigma_{\alpha\beta} q_1(0)]|0\rangle\,,
  \label{eq:correlationtensors}
  \end{eqnarray}
where $ \Pi_{\mu\nu\alpha\beta}$ can be decomposed into two
Lorentz invariant functions $\Pi^\pm$ as
\begin{eqnarray}
   \Pi_{\mu\nu\alpha\beta} &=&
   [g_{\mu\alpha}g_{\nu\beta} - g_{\mu\beta}g_{\nu\alpha}]\Pi^+(q^2)\
  \nonumber\\
  &&{}
  + [g_{\mu\beta} q_\nu q_\alpha
 +  g_{\nu\alpha} q_\mu q_\beta - g_{\mu\alpha} q_\nu q_\beta
  -  g_{\nu\beta} q_\mu q_\alpha] \frac{\Pi^+(q^2) + \Pi^-(q^2)}{q^2}\,.
  \label{eq:decompose}
  \end{eqnarray}
The $\Pi^{+,(-)}(q^2)$, corresponding to interpolating states with
positive (negative) parity, respectively,  were computed in
\cite{GRVW}:
  \begin{eqnarray}
   \Pi^{\mp}(q^2) &=& -\frac{1}{8\pi^2}q^2\ln\,\frac{-q^2}{\mu^2}\Bigg[
  1\pm 6\frac{m_1 m_2}{q^2}
  +\frac{\alpha_s}{3\pi}\Big(\ln\,\frac{-q^2}{\mu^2}+\frac{7}{3}\Big)\Bigg]
   \nonumber\\
   &&{}
  +\frac{1}{q^2} \Bigg[-\frac{1}{24 }\langle \frac{\alpha_s}{\pi}\,G^2\rangle
  + \Bigg( \frac{1}{2}m_1\pm m_2\Bigg)\langle \bar q_1 q_1\rangle
  + \Bigg( \frac{1}{2}m_2\pm m_1\Bigg)\langle \bar q_2
  q_2\rangle\Bigg]\nonumber\\
  &&{}
  -\frac{1}{q^4} \Bigg[ \frac{1}{6}\Bigg( 2 m_1\pm m_2\Bigg)
  \langle \bar q_1 g_s \sigma\cdot G q_1\rangle
  + \frac{1}{6}\Bigg( 2 m_2\pm m_1\Bigg)
  \langle \bar q_2 g_s \sigma\cdot G q_2\rangle\nonumber\\
  &&{}{}
  +  \frac{2}{9}\pi\alpha_s \bigg(
  \langle \bar q_1 \gamma_\mu \lambda^a q_1\,
   \bar q_1 \gamma^\mu \lambda^a q_1\rangle
  +\langle \bar q_2 \gamma_\mu \lambda^a q_2\,
  \bar q_2 \gamma^\mu \lambda^a q_2\rangle\bigg)\nonumber\\
  &&{}{}
  \pm 4 \pi \alpha_s
  \langle \bar q_1 \bigg\{\begin{array}{c}
    \gamma_5 \\
    1 \\
  \end{array}\bigg\} \lambda^a q_2\, \bar q_2 \bigg\{\begin{array}{c}
    \gamma_5 \\
    1 \\
  \end{array}\bigg\}\lambda^a q_1\rangle\Bigg]\,.
  \label{eq:pipm}
  \end{eqnarray}
While $\Pi^-(q^2)$ is relevant for extracting the value of
$f_V^\perp$~\cite{GRVW,Ball:1996tb}, one gets from $\Pi^+(q^2)$
the $f_A^\perp$ RG-improved sum rules:
  \begin{eqnarray}
  \lefteqn{
  m_{A}^2e^{-m_{A}^2/M^2} L^{-8/(3b)} (f_{A}^\perp)^2  }
  \nonumber\\
  &=&
  \frac{1}{8\pi^2}\int\limits_0^{s_0^{A}}\!\! sds\,e^{-s/M^2} \left(
  1 - 6\frac{m_1 m_2}{s}L^{-8/b}+ \frac{\alpha_s}{\pi}\left[\frac{7}{9}+\frac{2}{3}\,
  \ln\,\frac{s}{\mu^2}
  \right]\right)
  +\frac{1}{24}\,\langle\frac{\alpha_s}{\pi}G^2\rangle\nonumber\\
  &&
  -  \Bigg( \frac{1}{2}m_1 - m_2\Bigg)\langle \bar q_1 q_1\rangle
  L^{4/b}
  - \Bigg( \frac{1}{2}m_2 - m_1\Bigg)\langle \bar q_2
  q_2\rangle L^{4/b}\nonumber\\
  &&
  -\frac{1}{M^2} \Bigg[ \frac{1}{6} ( 2 m_1 - m_2)
  \langle \bar q_1 g_s \sigma\cdot G q_1\rangle  L^{-14/(3b)}
  + \frac{1}{6}( 2 m_2 - m_1)
  \langle \bar q_2 g_s \sigma\cdot G q_2\rangle L^{-14/(3b)}\nonumber\\
  &&\ \ \  -\frac{32\pi\alpha_s}{81}\,(\langle\bar q_1
  q_1\rangle^2 + \langle\bar q_2
  q_2\rangle^2) + \frac{16\pi\alpha_s}{9}\,\langle\bar q_1
  q_1\rangle \langle\bar q_2
  q_2\rangle \Bigg]\,,\makebox[0.8cm]{}
  \label{eq:SRft3}
  \end{eqnarray}
where $L\equiv \alpha_s(\mu)/\alpha_s(M)$, $M$ is the Borel mass,
$s_0^{A}$ is the threshold for higher resonances, and $f^\perp_A$
depends on the renormalization scale as
 \begin{eqnarray}
 f^\perp_A(\mu) =f^\perp_A(\mu_0)
 \left(\frac{\alpha_s(\mu_0)}{\alpha_s(\mu)}\right)^{-{4\over
 3b}}.
 \end{eqnarray}

We then start with the analysis of the $f_A^\perp$ sum rules. We
take the experimental mass results for $b_1$ and $h_1$ but with
larger uncertainties: $m_{b_1}=(1300\pm 70)$~MeV,
$m_{h_1}=(1386\pm70)$~MeV as inputs to be consistent with QCD sum
rule calculations~\cite{GRVW}. We also consider the isodoublet
strange meson $K_{1B}$ of quantum number $1^1P_1$. It should be
noted that the real physical states $K_1(1270)$ and $K_1(1400)$
are the mixture of $1^3P_1$ and $1^1P_1$ states. Following the
notations in Ref.~\cite{Suzuki:1993yc}, the relations can be
written as
 \begin{eqnarray}
 \label{eq:mixing}
 K_1(1270) &=& K_{1A}\sin\theta_K+K_{1B}\cos\theta_K, \nonumber \\
 K_1(1400) &=& K_{1A}\cos\theta_K-K_{1B}\sin\theta_K,
 \end{eqnarray}
where $K_{1A}$ is the strange mesons of quantum numbers $1^3P_1$.
The mixing angle $\theta_K$ may be close to $45^\circ$~\cite{PDG}.
In the following study, we take $m_{K_{1B}}=1370\pm
70$~MeV~\cite{GRVW}, where the uncertainty is also enlarged. We
obtain the tensor couplings (at scale 1~GeV),
 \begin{eqnarray}\label{eq:tensorcoupling}
 && f_{b_1}^\perp=(180\pm 10)~{\rm MeV} \ \ {\rm for}\ \ s_0^{b_1}=(2.6\pm 0.3)~{\rm GeV}^2  ,
 \nonumber\\
 && f_{h_1}^\perp=(200\pm 20)~{\rm MeV} \ \ {\rm for}\ \ s_0^{h_1}=(3.5\pm 0.3)~{\rm GeV}^2  ,
 \nonumber\\
 && f_{K_{1B}}^\perp=(195\pm 10)~{\rm MeV} \ \ {\rm for}\ \ s_0^{K_{1B}}=(3.1\pm 0.3)~{\rm GeV}^2   ,
 \end{eqnarray}
where the values of $s_0$ are determined when the stability of the
sum rules is reached within the Borel window
1~GeV$^2<M^2<1.5$~GeV$^2$. Note that $K_{1A}$ couples to the local
axial vector current, instead of the local tensor current; in
other words, one has
 $\langle 0 |\bar q\gamma_\mu\gamma_5 s |K_{1A}(P,\lambda)\rangle
 = -i f_{K_{1A}} \, m_{K_{1A}}\epsilon^{(\lambda)}_\mu$.
Therefore, according to Eq. (\ref{eq:mixing}), we have $\langle 0
|\bar q\sigma_{\mu\nu}s |K_{1}(1270)(P,\lambda)\rangle$ $= i
f_{K_{1B}}^\perp \cos {\theta_K}\,\epsilon_{\mu\nu\alpha\beta}
\epsilon_{(\lambda)}^\alpha P^\beta$ and $\langle 0 |\bar
q\sigma_{\mu\nu}s |K_{1}(1400)(P,\lambda)\rangle$ $= -i
f_{K_{1B}}^\perp \sin {\theta_K}\,\epsilon_{\mu\nu\alpha\beta}
\epsilon_{(\lambda)}^\alpha P^\beta$ with $\bar q\equiv \bar u,
\bar d$.

\subsection{The Gegenbauer moments for LCDAs
$\Phi_\parallel^A$}\label{subsec:moments_even}

\begin{figure}[ht]
\epsfxsize=3in \centerline{\epsffile{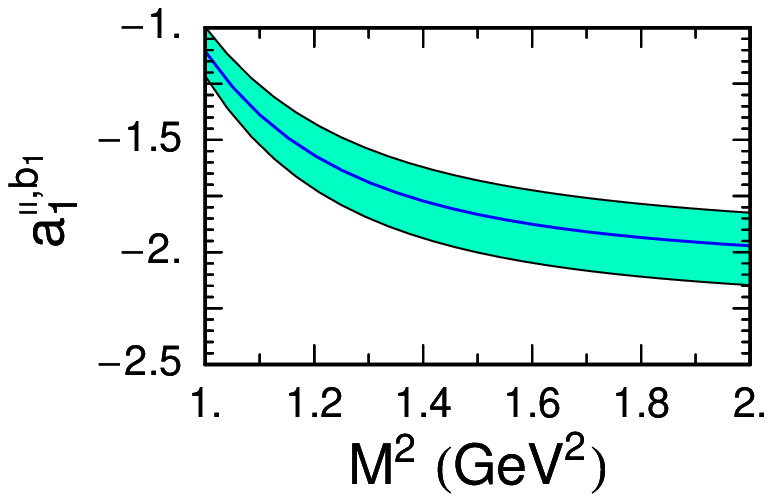}}
\centerline{\parbox{14cm}{\caption{\label{fig:a1pb1}
$a_1^{\parallel, b_1}(1~{\rm GeV})$ as a function of the Borel
mass squared for $f_{b_1}=f_{b_1}^\perp(1~{\rm GeV})=180~{\rm
MeV}$. The band corresponds to the uncertainties of the input
parameters.}}}
\end{figure}

\begin{figure}[ht]
\epsfxsize=3in \centerline{\epsffile{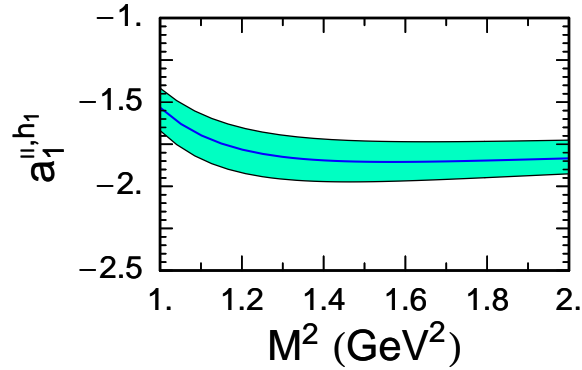}}
\centerline{\parbox{14cm}{\caption{\label{fig:a1ph1}
$a_1^{\parallel, h_1}(1~{\rm GeV})$ with
$f_{h_1}=f_{h_1}^\perp(1~{\rm GeV})=200~{\rm MeV}$. Others are the
same as Fig.~\ref{fig:a1pb1}. }}}
\end{figure}

\begin{figure}[ht]
\epsfxsize=3in \centerline{\epsffile{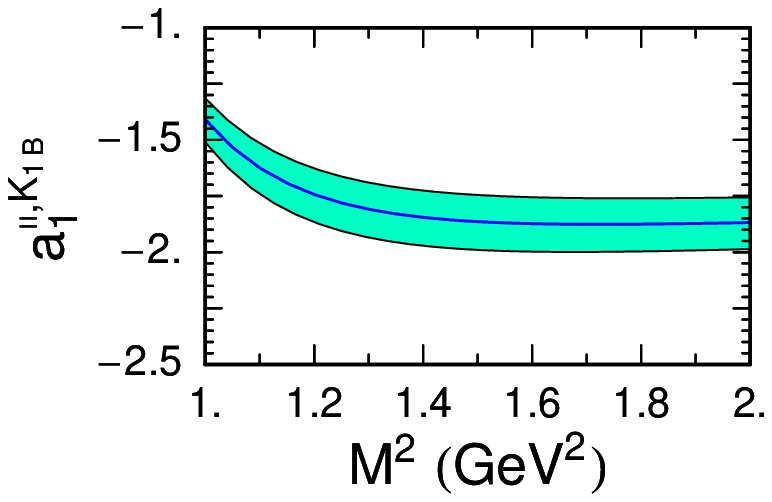}}
\centerline{\parbox{14cm}{\caption{\label{fig:a1pK1B}
$a_1^{\parallel, K_{1B}}(1~{\rm GeV})$ with
$f_{K_{1B}}=f_{K_{1B}}^\perp(1~{\rm GeV})=195~{\rm MeV}$. Others
are the same as Fig.~\ref{fig:a1pb1}. }}}
\end{figure}

\begin{figure}[th]
\begin{center}
 \centerline{
{\epsfxsize2.7in \epsffile{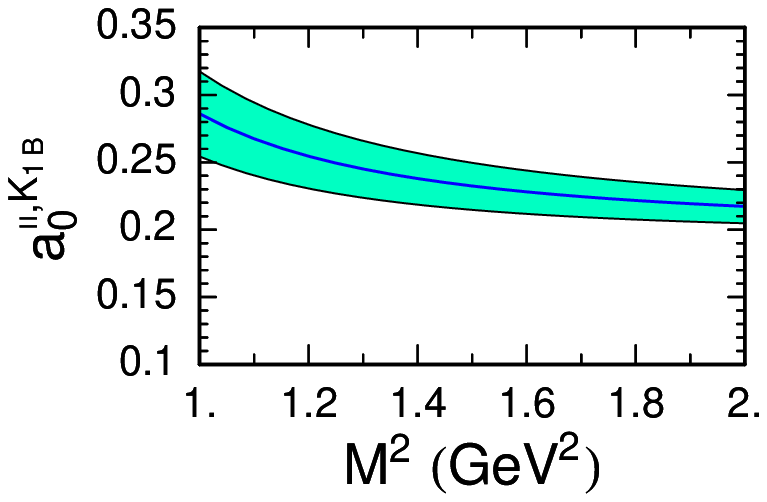}} \hspace{0.5cm}
{\epsfxsize2.7in \epsffile{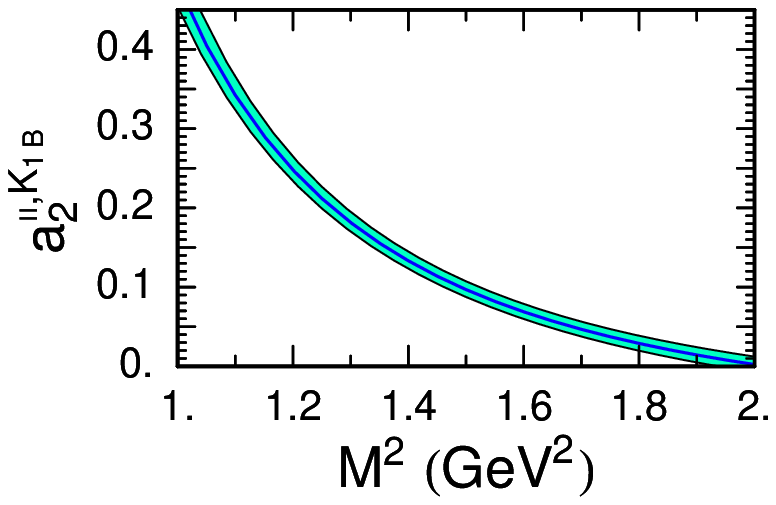}} }
\centerline{\parbox{14cm}{\caption{\label{fig:aevenpK1B}
$a_{0,2}^{\parallel, K_{1B}}(1~{\rm GeV})$ with
$f_{K_{1B}}=f_{K_{1B}}^\perp(1~{\rm GeV})=195~{\rm MeV}$. Others
are the same as Fig.~\ref{fig:a1pb1}. }}}
\end{center}
\end{figure}

The LCDAs $\Phi_\parallel^A(u,\mu)$ corresponding to the $1^1P_1$
states are defined as\footnote{\label{footnote:K1B} It is known
that the coupling of the $K_{1B}$ to the local axial-vector
current does not vanish in the isospin limit:
 \begin{equation}
 \langle K_{1B}(P,\lambda)|\bar s(0)\gamma_\mu\gamma_5 q(0)|0\rangle
 =i\bar f_{K_{1B}} m_{K_{1B}} \epsilon_\mu^{*(\lambda)}, \nonumber
  \end{equation}
i.e., $\bar f_{K_{1B}}\not=0$. Thus one can define the chiral-even
leading-twist LCDA  as
\begin{equation} \langle K_{1B}(P,\lambda)|\bar
s(y)\not\! z\gamma_5 q(x)|0\rangle = i \bar f_{K_{1B}} m_{K_{1B}}
(\epsilon_\lambda^*\cdot z)\int^1_0 dx e^{i(up\cdot y +\bar
up\cdot x)}\bar\Phi_\parallel^{K_{1B}} (u,\mu),
 \end{equation}
 where
\begin{eqnarray}
\bar\Phi_\parallel^{K_{1B}}(u,\mu)=6u(1-u) \Bigg(1 +
\sum_{l=1}^\infty \bar a_l^{\parallel}(\mu) C^{3/2}_l(2u-1)
\Bigg).
\end{eqnarray}
Essentially, the above definition is consistent with what we
adopted in the present paper, satisfying the following relations
 \begin{eqnarray}
 a_0^\parallel f_{K_{1B}} =\bar f_{K_{1B}}, \hspace{0.5cm}
 a_l^\parallel f_{K_{1B}} =\bar a_l^\parallel \bar f_{K_{1B}}  {~~\rm for}~l\geq
 1,
 \end{eqnarray}
and
\begin{eqnarray}
 \int_0^1 \Phi_\parallel^{K_{1B}}(u,\mu)du = a_0^\parallel, \hspace{0.5cm}
 \int_0^1 \bar\Phi_\parallel^{K_{1B}}(u,\mu) du=1.
 \end{eqnarray}
 }
 \begin{equation}
\langle A(P,\lambda)|\bar q_1(y)\not\! z\gamma_5 q_2(x)|0\rangle =
i f_A m_A (\epsilon_\lambda^*\cdot z)\int^1_0 dx e^{i(up\cdot y
+\bar up\cdot x)}\Phi_\parallel^A(u,\mu),
 \end{equation}
where $z^2=(y-x)^2=0$. $\Phi_\parallel^A(u,\mu)$ can be expanded
in a series of Gegenbauer polynomials as given in
Eq.~(\ref{eq:generaldaparallel}). To calculate the Gegenbauer
moments of $\Phi_\parallel^A$, we consider the following two-point
correlation functions
\begin{eqnarray}\label{eq:2pointaparallel}
\Pi_{l,\mu\nu}  = i\int d^4x e^{iqx} \langle 0| T(\Omega^A_l(x)\
O_{\mu\nu}^\dagger(0) |0 \rangle = (zq)^{l} I_l (q^2) (z_\mu q_\nu
- z_\nu q_\mu),
\end{eqnarray}
where, to leading logarithmic accuracy, the relevant
multiplicatively renormalizable operator is
  \begin{eqnarray}
  \Omega^A_{l}(x) &=& \sum\limits_{j=0}^l c_{l,j} (iz\partial)^{l-j}
  \bar{q}_2(x)\!\not\!z \gamma_5\,(iz\deriv)^j q_1(x)\,,
  \label{eq:optensor}
\end{eqnarray}
with $\deriv_\mu = \derright_\mu - \derleft_\mu =
(\stackrel{\rightarrow}{\partial} +ig_s A^a(x)\lambda^a/2)_\mu-
(\stackrel{\leftarrow}{\partial} -ig_s A^a(x)\lambda^a/2)_\mu$,
$c_{l,k}$ being the coefficients of the Gegenbauer polynomials
such that $C_l^{3/2}(x) = \sum c_{l,k}x^k$. $\Omega_l^A$ and
$O_{\mu\nu}$ satisfy the following relations:

\begin{eqnarray}
\!\! \langle 0| \Omega^A_l(0)|A(P,\lambda)\rangle &=& -if_A
m_A(\epsilon_\lambda \cdot z)(z\cdot P)^l
\frac{3(l+1)(l+2)}{2(2l+3)}\, a_l^{\parallel,A}(\mu)\,,
\\
\langle 0| O_{\mu\nu}(0)|A(P,\lambda)\rangle &\equiv& \langle 0|
\bar q_2(0) i\sigma_{\mu\nu}\gamma_5 q_1(0) |A(P,\lambda)\rangle =
-i f_A^\perp (\epsilon^{(\lambda)}_\mu P_\nu
-\epsilon^{(\lambda)}_\nu P_\mu).\hspace{1.2cm}
\end{eqnarray}
The RG-improved sum rules for Gegenbauer moments
$a_l^{\parallel,A}$ read
\begin{eqnarray}\label{eq:srgmomentsparallel}
a_l^{\parallel,A} &&= -\frac{1}{m_A f_A f_A^\perp}
e^{m_A^2/M^2} \frac{2(2l+3)}{3(l+1)(l+2)} L^{[(4/3)+\gamma_{(l)}^\parallel]/b}\nonumber\\
&&\times \Bigg\{ \frac{3 }{4\pi^2 }M^2
(1-e^{-s_0^{\parallel,A}/M^2})
 \Bigg( \int^1_0 d\alpha C_l^{3/2}(2\alpha -1) [m_{q_2} \alpha + m_{q_1}
 (\alpha-1)] \Bigg)L^{-4/b}\nonumber\\
 && - C_l^{3/2}(1)(\langle \bar q_2 q_2\rangle + \langle \bar q_1 q_1\rangle
  (-1)^{l+1}) L^{4/b}\nonumber\\
 && - \Bigg( \frac{1}{3} C_l^{3/2}(1) + 2 C_{l-1}^{5/2}(1)\theta(l-1) \Bigg)
   \frac{\langle \bar q_2 g_s \sigma\cdot G q_2\rangle
   + \langle \bar q_1 g_s \sigma\cdot G q_1
 \rangle (-1)^{l+1}}{M^2}L^{-2/(3b)}\nonumber\\
 && - \frac{\pi^2}{M^4}
  \Big[\frac{20}{3} C_{l-2}^{7/2}(1)\theta(l-2)
  +\frac{1}{3} C_{l-1}^{5/2}(1)\theta(l-1))\Big]\nonumber\\
  && \ \ \times \langle \frac{\alpha_s}{\pi} G^2\rangle
 [\langle \bar q_2 q_2\rangle + \langle \bar q_1 q_1\rangle
  (-1)^{l+1}]L^{4/b} \Bigg\}.
 \end{eqnarray}
 The above sum rules for Gegenbauer moments $a_l^{\parallel,A}$
 amount to the results in terms of moments $\langle
\xi_{\parallel,A}^l \rangle$:
\begin{eqnarray}\label{eq:srmoments}
\langle \xi^l_{\parallel,A}\rangle
 &=& -\frac{1}{m_A f_A f_A^\perp} e^{m_A^2/M^2}\nonumber\\
&&\times \Bigg\{ \Bigg[ \frac{3 }{16\pi^2 }M^2 \Bigg(\frac{m_{q_2}
+ m_{q_1}}{l+2} + \frac{m_{q_2} - m_{q_1}}{l+1} \Bigg) - \langle
 \bar q_2 q_2\rangle \nonumber\\
 && - \frac{2l+1}{3} \frac{\langle \bar q_2 g_s \sigma\cdot G q_2
 \rangle}{M^2} -
 \frac{\pi^2 l}{9 M^4} (4l-3)\langle\frac{\alpha_s}{\pi} G^2\rangle
  \langle \bar q_2 q_2\rangle
 \Bigg]
 \nonumber\\
 && +  (-1)^{l+1}  \Bigg[ \frac{3 }{16\pi^2 }M^2 \Bigg(\frac{m_{q_2}
+ m_{q_1}}{l+2} - \frac{m_{q_2} - m_{q_1}}{l+1} \Bigg)  - \langle
 \bar q_1 q_1\rangle \nonumber\\
 && -\frac{2l+1}{3} \frac{\langle \bar q_1 g_s \sigma\cdot G q_1
\rangle}{M^2} -
 \frac{\pi^2 l}{9 M^4} (4l-3)\langle\frac{\alpha_s}{\pi} G^2\rangle
  \langle \bar q_1 q_1\rangle
 \Bigg]\Bigg\},
 \end{eqnarray}
 where
 \begin{eqnarray}
 \langle \xi^l_{\parallel,A}\rangle_\mu = \int_0^1 dx\, (2x-1)
\Phi_\parallel^A(x,\mu).
\end{eqnarray}
Five remarks are in order. First, we will simply take $f_{A}
=f_A^\perp(\mu=1$~GeV) in the study since only the products of
$f_A a_l^{\parallel,A}$ are relevant. Second, the sum rules
obtained from the nondiagonal correlation functions in
Eq.~(\ref{eq:2pointaparallel}) can also determine the sign of $f_A
a_l^{\parallel, A}$ relative to $f_A^\perp$. Third, the sum rules
for $\langle \xi_{\parallel,A}^l \rangle$ cannot be improved by RG
equation since $\langle \xi_{\parallel,A}^l \rangle$ mix with each
other even in the one-loop level. For the present case, the RG
effects are relatively small compared with the uncertainties of
input parameters. Fourth, neglecting the small isospin violation,
$a_0^\parallel$ and $a_2^\parallel$ are nonzero only for $K_{1B}$.
Fifth, in the large $l$ limit, the actual expansion parameter is
$M^2/l$ for moment sum rules. One can find that, for $l\geq 3$ and
fixed $M^2$, the operator-product-expansion (OPE) series may
become divergent. In other words, it is impossible to obtain
reliable $a_l^{\parallel, A}$ with $l\geq 3$. In the numerical
analysis, we therefore choose the Borel window $(1.0+l)$~GeV$^2<
M^2 < 2.0$~GeV$^2$ for $a_l^{\parallel,}$ with $l\leq 2$, where
the contributions originating from higher resonances and the
highest OPE terms are well under control. Using the input
parameters given in Sec.~\ref{subsec:inputs}, we  obtain the
Gegenbauer moments, corresponding to $\mu_1\equiv 1$~GeV and
$\mu_2\equiv 2.2$~GeV,
 \begin{eqnarray}\label{eq:gvalues1}
 & & a_1^{\parallel, b_1}(\mu_{1[2]})=(-1.70\pm 0.45)
\frac{(180~{\rm MeV})^2}{f_{b_1}\, f_{b_1}^\perp(\mu_1)}  \,,
   \
  [{\rm or}]\ (-1.41\pm 0.37)
 \frac{(180)(165)\,{\rm MeV}^2}{f_{b_1}\, f_{b_1}^\perp(\mu_2)}\,,\nonumber\\
 & & a_1^{\parallel, h_1}(\mu_{1[2]})=(-1.75\pm 0.20)
 \frac{(200~{\rm MeV})^2}{f_{h_1}\, f_{h_1}^\perp(\mu_1)}
 \,,  \
  [{\rm or}]\ (-1.45\pm 0.17)
   \frac{(200)(183)\,{\rm MeV}^2}{f_{h_1}\, f_{h_1}^\perp(\mu_2)}\,,\nonumber\\
 & & a_1^{\parallel, K_{1B}}(\mu_{1[2]})=(-1.75\pm 0.25)
 \frac{(195~{\rm MeV})^2}{f_{K_{1B}}\, f_{K_{1B}}^\perp(\mu_1)},
  [{\rm or}]\ (-1.45\pm 0.21)
  \frac{(195)(179)\,{\rm MeV}^2}{f_{K_{1B}}\,
  f_{K_{1B}}^\perp(\mu_2)},
  \nonumber\\
  & & a_0^{\parallel, K_{1B}}(\mu_{1[2]})=(0.26\pm 0.06)
  \frac{(195~{\rm MeV})^2}{f_{K_{1B}}\, f_{K_{1B}}^\perp(\mu_1)}
  \,,  \
  [{\rm or}]\ (0.26\pm 0.06)
  \frac{(195)(179)\,{\rm MeV}^2}{f_{K_{1B}}\,
   f_{K_{1B}}^\perp(\mu_2)}, \nonumber \\
 & & a_2^{\parallel, K_{1B}}(\mu_{1[2]})=(0.13\pm
  0.13) \frac{(195~{\rm MeV})^2}{f_{K_{1B}}\, f_{K_{1B}}^\perp(\mu_1)}\,,
  \
  [{\rm or}]\ (0.10\pm 0.10)
   \frac{(195)(179)\,{\rm MeV}^2}{f_{K_{1B}}\,
   f_{K_{1B}}^\perp(\mu_2)}\,.\nonumber\\
 \end{eqnarray}
The Gegenbauer moments versus the Borel Mass squared are plotted
in Figs.~\ref{fig:a1pb1}-\ref{fig:aevenpK1B}.


\subsection{The Gegenbauer moments for LCDAs $\Phi_\perp^A$}\label{subsec:moments_odd}

To disentangle the contribution of $\Phi_\perp^A$ from higher
twist DAs,  it is unavoidable to have an admixture of negative
parity vector states in the QCD sum rule study. We consider the
following correlation functions
\begin{eqnarray}
\Pi_l (q^2, qz) = i\int d^4x e^{iqx} \langle 0| T(\Omega^{A,T\,
\mu}_l(x)\ O_{\mu}^\dagger(0) |0 \rangle =
  2(qz)^{l+2}I_l^{\perp}(q^2),
  \label{eq:amoment2perp}
\end{eqnarray}
where, to leading logarithmic accuracy, the relevant
multiplicatively renormalizable operator is
  \begin{eqnarray}
  \Omega^{A,T\, \mu}_{l}(x) &=& \sum\limits_{j=0}^l c_{l,j} (iz\partial)^{l-j}
  \bar{q}_2(x)\!\sigma^{\mu\alpha}\gamma_5 z_\alpha \,(iz\deriv)^j q_1(x)\,,
  \label{eq:optensor1}
\end{eqnarray}
and
\begin{eqnarray}
  O_{\mu}(0) &=& \bar q_1(0)\sigma_{\mu\beta}\gamma_5 z^\beta q_2(0)\,,
  \label{eq:optensor2}
\end{eqnarray}
which satisfy the following relations:

\begin{eqnarray}
 \sum_\lambda \langle 0| \Omega^{A,T\, \mu}_{l}(0)|A(P,\lambda)\rangle
 \langle A(P,\lambda)| O_{\mu}(0)|0 \rangle
  &=& 2 (f_A^\perp)^2 (z P)^{l+2}
\frac{3(l+1)(l+2)}{2(2l+3)}\, a_l^{\perp,A}\,, \nonumber\\
 \sum_\lambda \langle 0| \Omega^{A,T\, \mu}_{l}(0)|V(P,\lambda)\rangle
 \langle V(P,\lambda)| O_{\mu}(0)|0 \rangle
  &=& 2 (f_V^\perp)^2 (z P)^{l+2}
\frac{3(l+1)(l+2)}{2(2l+3)}\, a_l^{\perp,V}\,.\nonumber\\
\end{eqnarray}
Here $V$ refers to the vector mesons. The Gegenbauer moment sum
rules read (c.f. Ref.~\cite{Ball:2003sc})

\begin{eqnarray}
 a_l^{\perp,A} &=& \frac{1}{ (f_{A}^{\perp})^2}
e^{m_{A}^2/M^2} \frac{2(2l+3)}{3(l+1)(l+2)} \nonumber\\
&&\times \Bigg\{\frac{1}{2\pi^2}\,\frac{\alpha_s}{\pi}\, M^2
\left(1-e^{-s_0^{\perp,A}/M^2} \right) \int_0^1\!\! du\, u\bar u\,
C_l^{3/2}(2u-1) \left( \ln u + \ln \bar u + \ln^2\,\frac{u}{\bar
u}\right)
\nonumber\\
& & {} + \frac{1}{12 M^2}\, \langle \frac{\alpha_s}{\pi}
G^2\rangle \bigg(C_l^{3/2}(1)-2 \bigg)\nonumber\\
&& +
\frac{1}{M^2}\,
 (m_{q_1} \langle \bar q_1 q_1\rangle + m_{q_2} \langle \bar q_2 q_2\rangle)
C_l^{3/2}(1)\nonumber\\
&&  + \frac{1}{3 M^4}\,
 (m_{q_1}\langle \bar q_1 g_s\sigma G q_1\rangle
 + m_{q_2}\langle \bar q_2 g_s\sigma G q_2\rangle)
 \bigg( 3 C_{l-1}^{5/2}(1)\theta(l-1) + C_l^{3/2}(1) \bigg)
 \nonumber\\
& & {} + \frac{32\pi\alpha_s}{81M^4}\,
 (\langle \bar q_1 q_1\rangle^2+ \langle \bar q_2 q_2\rangle^2)
\bigg( 3 C_{l-1}^{5/2}(1)\theta(l-1) - C_l^{3/2}(1) \bigg)\nonumber\\
&& -\frac{3(l+1)(l+2)}{2(2l+3)}\,(f_{V}^\perp)^2 a_l^{\perp,V}
e^{-m_{V}^2/M^2} \Bigg\}
\end{eqnarray}
for even $l\geq 2$, and
\begin{eqnarray}\label{eq:srodd}
 a_l^{\perp,A} &=& \frac{1}{ (f_{A}^{\perp})^2}
e^{m_{A}^2/M^2} \frac{2(2l+3)}{3(l+1)(l+2)} \nonumber\\
&&\times \Bigg\{- \frac{1}{M^2}\, (m_{q_1} \langle \bar q_1
q_1\rangle - m_{q_2}
 \langle \bar q_2 q_2\rangle)
C_l^{3/2}(1)\nonumber\\
&& - \frac{1}{3M^4}\, (m_{q_1}\langle \bar q_1
 g_s\sigma G q_1\rangle - m_{q_2}\langle \bar q_2 g_s\sigma G
 q_2\rangle)
\bigg( 3 C_{l-1}^{5/2}(1)\theta(l-1) + C_l^{3/2}(1) \bigg)\nonumber\\
& & {} - \frac{32\pi\alpha_s}{81M^4}\,
 (\langle \bar q_1 q_1\rangle^2 - \langle \bar q_2 q_2 \rangle^2)
 \bigg( 3 C_{l-1}^{5/2}(1)\theta(l-1) - C_l^{3/2}(1) \bigg)
 \nonumber\\
&& -\frac{3(l+1)(l+2)}{2(2l+3)}\,(f_{V}^\perp)^2 a_l^{\perp,V}
e^{-m_{V}^2/M^2}\Bigg\}
\end{eqnarray}
for odd $l$.

\begin{figure}[th]
\epsfxsize=3in \centerline{\epsffile{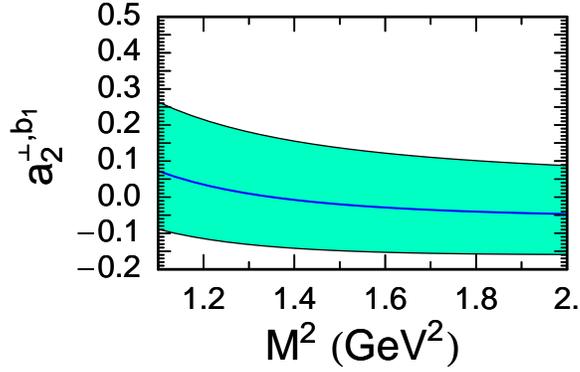}}
\centerline{\parbox{14cm}{\caption{\label{fig:a2perpb1}
$a_2^{\perp, b_1}(1~{\rm GeV})$ as a function of the Borel mass
squared for $f_{b_1}^\perp(1~{\rm GeV})=180~{\rm MeV}$. The band
corresponds to the uncertainties of the input parameters.}}}
\end{figure}

\begin{figure}[th]
\epsfxsize=3in \centerline{\epsffile{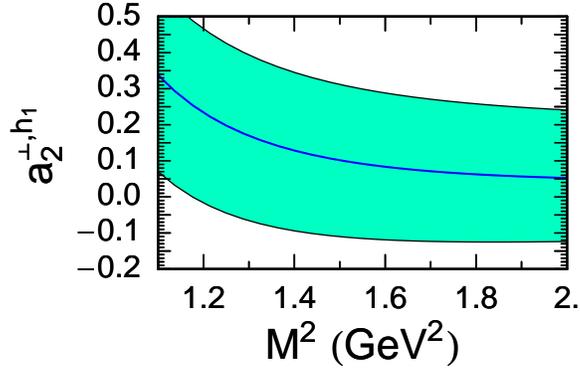}}
\centerline{\parbox{14cm}{\caption{\label{fig:a2perph1}
$a_2^{\perp, h_1}(1~{\rm GeV})$ with $f_{h_1}^\perp(1~{\rm
GeV})=200~{\rm MeV}$. Others are the same as
Fig.~\ref{fig:a2perpb1}.}}}
\end{figure}

\begin{figure}[th]
\epsfxsize=3in \centerline{\epsffile{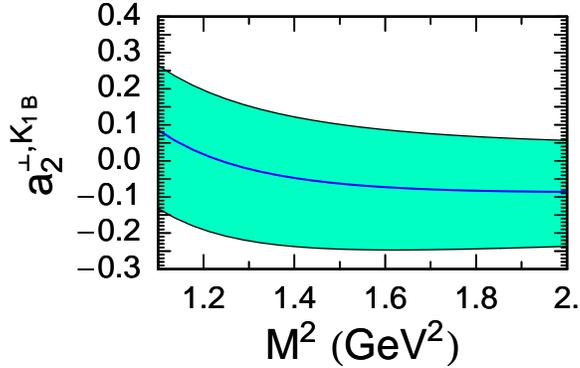}}
\centerline{\parbox{14cm}{\caption{\label{fig:a2perpK1B}
$a_2^{\perp, K_{1B}}(1~{\rm GeV})$  with $f_{K_{1B}}^\perp(1~{\rm
GeV})=195~{\rm MeV}$. Others are the same as
Fig.~\ref{fig:a2perpb1}.}}}
\end{figure}

\begin{figure}[th]
\begin{center}
 \centerline{
{\epsfxsize2.7 in \epsffile{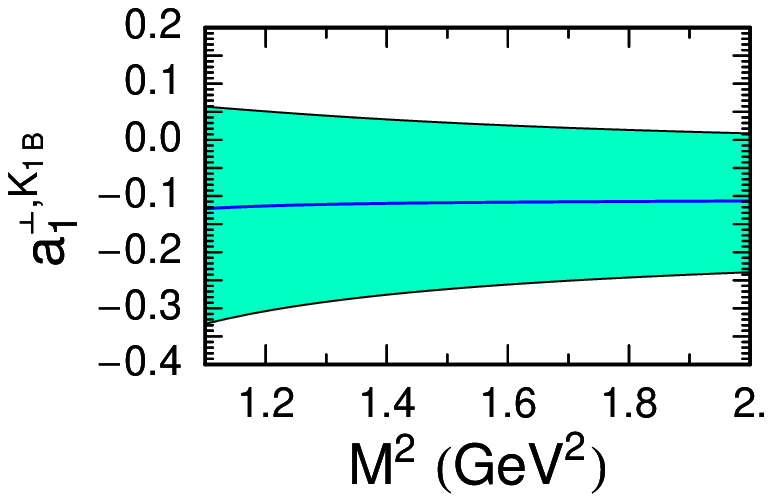}} \hspace{0.5cm}
{\epsfxsize2.7 in \epsffile{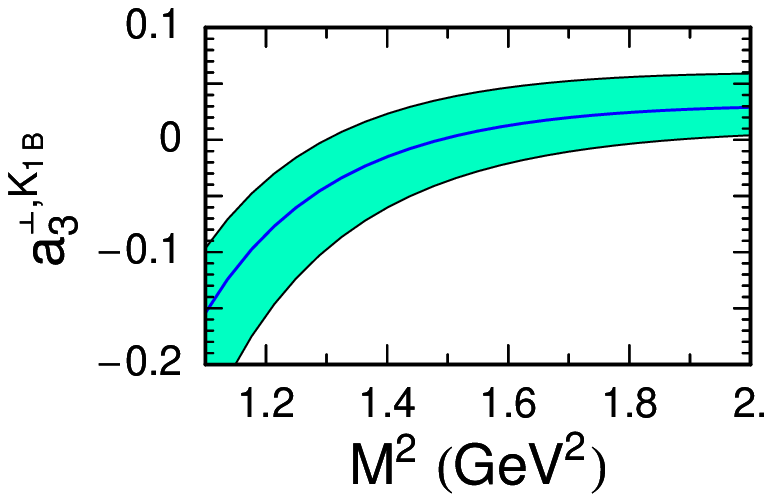}}
            }
\centerline{\parbox{14cm}{\caption{\label{fig:a13perpK1B}
$a_1^{\perp, K_{1B}}(1~{\rm GeV})$ and $a_3^{\perp, K_{1B}}(1~{\rm
GeV})$  with  $f_{K_{1B}}^\perp(1~{\rm GeV})=195~{\rm MeV}$.
Others are the same as Fig.~\ref{fig:a2perpb1}.}}}
\end{center}
\end{figure}

It should be noted again that for the present Gegenbauer moment
sum rules  the actual expansion parameter is $M^2/l$ in the large
$l$ limit. As a result, for $l \geq 4$ and fixed $M^2$, the OPE
series are convergent slowly or even divergent. In the numerical
analysis, we choose the Borel windows (i) $1.1$~GeV$^2< M^2 <
2.0$~GeV$^2$ for $a_1^{\perp, K_{1B}}$, (ii) $1.2$~GeV$^2< M^2 <
2.0$~GeV$^2$ for $a_2^{\perp, A}$, and (iii) $1.3$~GeV$^2< M^2 <
2.0$~GeV$^2$ for $a_3^{\perp, K_{1B}}$ where the contributions
originating from higher resonances and the highest OPE terms are
under control. We use the parameters given in
Sec.~\ref{subsec:inputs} and the relevant Gegenbauer moments of
the LCDAs of vector mesons as inputs. $a_2^{\perp, V}$ at scale
1~GeV are summarized as below~\cite{Ball:2004rg}:
\begin{eqnarray}\label{eq:gmomentsvector}
a_2^{\perp,\rho} &=& 0.2\pm 0.1,\nonumber\\
a_2^{\perp,\phi} &=& 0.0\pm 0.1, \nonumber\\
a_2^{\perp,K^*} &=& 0.13\pm 0.08.
\end{eqnarray}
$a_1^{\perp, V}$, which is obvious nonzero only for $K^*$, was
studied in Ref.~\cite{Ball:2003sc}, where a sign error was found
as compared to the original work of Chernyak and
Zhitnitsky~\cite{CZreport}. The authors in Ref.~\cite{Ball:2003sc}
thus concluded that $a_1$, which refers to a $K^{(*)}$ containing
an $s$ quark\footnote{$a_1$ changes sign for a $K^{(*)}$ involving
an $\bar s$ quark. In the present paper, we adopt the convention
for $a_1$ referring to $K^{(*)}$ and $K_{1B}$ of containing an $s$
quark.}, should be negative. Nevertheless, Braun and
Lenz~\cite{Braun:2004vf} have analyzed $a_1^{\parallel,K^*}$ and
found the result to be $0.10 \pm 0.07$. They also argued that the
sum rule results in Ref.~\cite{Ball:2003sc} for
$a_1^{\parallel,K^*}$, $a_1^{\perp,K^*}$ and $a_1^{K}$ may be
unstable owing to many cancellations among OPE terms. Here, if
neglecting the $K_{1B}$ effect, we obtain $a_1^{\perp,K^*}=0.05\pm
0.02$ and $a_3^{\perp,K^*}= 0.02\pm 0.02$, where the value of
$a_1^{\perp,K^*}$ is consistent with that in
Ref.~\cite{Braun:2004vf}. In the present study, we will use
$a_1^{\perp,K^*}\simeq a_1^{\parallel,K^*} =0.10\pm
0.07$~\cite{Braun:2004vf,Ball:2004rg} and $a_3^{\perp,K^*}=
0.02\pm 0.02$ at the scale 1~GeV. The numerical analysis yields
 \begin{eqnarray}\label{eq:gvalues2}
 & & a_2^{\perp, b_1}(\mu_{1[2]})=(0.03\pm 0.19)
     \Bigg(\frac{180~{\rm MeV}}{f_{b_1}^\perp(\mu_1)} \Bigg)^2 ,
  \ \ \ \ \
  [{\rm or}]\ (0.02\pm 0.15)
 \Bigg(\frac{165~{\rm MeV}}{f_{b_1}^\perp (\mu_2)}\Bigg)^2, \nonumber\\
 & & a_2^{\perp,h_1}(\mu_{1[2]})=(0.17\pm 0.29)
  \Bigg(\frac{200~{\rm MeV}}{f_{h_1}^\perp(\mu_1)} \Bigg)^2 ,
  \ \ \ \ \
  [{\rm or}]\ (0.13\pm 0.23)
  \Bigg(\frac{183~{\rm MeV}}{f_{h_1}^\perp(\mu_2)} \Bigg)^2,\nonumber\\
 & & a_2^{\perp, K_{1B}}(\mu_{1[2]})=(-0.02\pm 0.22)
 \Bigg(\frac{195~{\rm MeV}}{f_{K_{1B}}^\perp (\mu_1)} \Bigg)^2,
   \
  [{\rm or}]\ =(-0.02\pm 0.17)
  \Bigg(\frac{179~{\rm MeV}}{f_{K_{1B}}^\perp (\mu_2)}\Bigg)^2,
  \nonumber\\
& & a_1^{\perp, K_{1B}}(\mu_{1[2]})=(-0.13\pm 0.19)
 \Bigg(\frac{195~{\rm MeV}}{f_{K_{1B}}^\perp (\mu_1)} \Bigg)^2,
  \
  [{\rm or}]\ (-0.11\pm 0.17)
  \Bigg(\frac{179~{\rm MeV}}{f_{K_{1B}}^\perp (\mu_2)}\Bigg)^2, \nonumber\\
  & & a_3^{\perp, K_{1B}}{(\mu_{1[2]})}=(-0.02\pm 0.08)
 \Bigg(\frac{195~{\rm MeV}}{f_{K_{1B}}^\perp (\mu_1)} \Bigg)^2,
 \
  [{\rm or}]\ (-0.01\pm 0.06)
  \Bigg(\frac{179~{\rm MeV}}{f_{K_{1B}}^\perp (\mu_2)}
  \Bigg)^2, \nonumber\\
 \end{eqnarray}
corresponding to the excited state thresholds
$s_0^{\perp,b_1}=2.3\pm 0.3~{\rm GeV}^2, s_0^{\perp,h_1}=2.7\pm
0.3~{\rm GeV}^2$ and $s_0^{\perp,K_{1B}}=2.5\pm 0.3~{\rm GeV}^2$,
respectively, where $\mu_1\equiv 1$~GeV and $\mu_2\equiv 2.2$~GeV.
To consider the vector modes in the excited states, lower
magnitudes of thresholds are taken here. The results for
Gegenbauer moments are insensitive to the thresholds. In
Figs.~\ref{fig:a2perpb1}-\ref{fig:a13perpK1B} we plot the
Gegenbauer moments as functions of the Borel mass squared, where
main uncertainties come from the errors of $a_l^{\perp, V}$.


\section{Summary}
\begin{figure}[t]
\begin{center}
 \centerline{
{\epsfxsize2.7in \epsffile{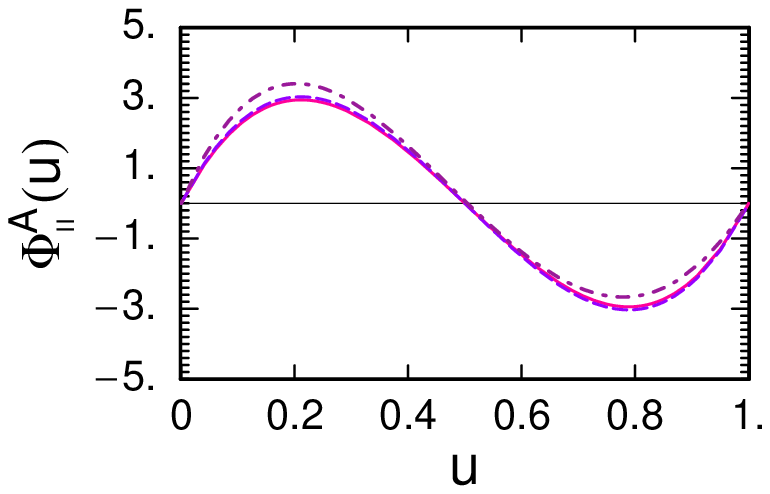}} \hspace{0.5cm}
{\epsfxsize2.8in \epsffile{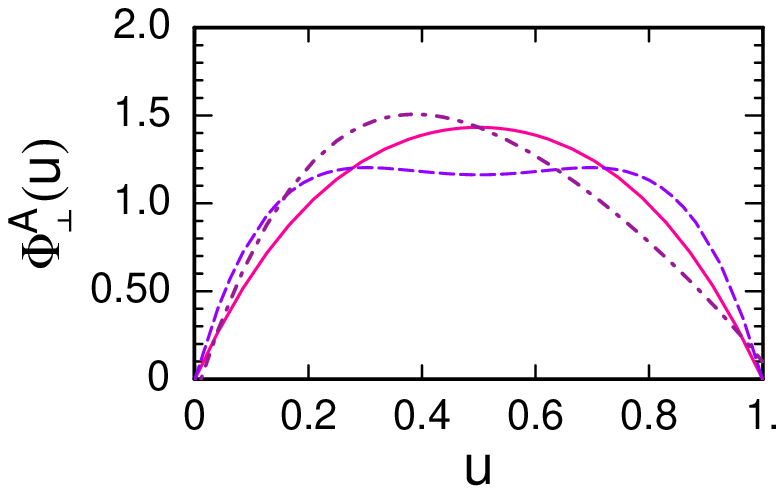}}
            }
\centerline{\parbox{14cm}{\caption{\label{fig:lcdas} Leading twist
light-cone distribution amplitudes at the scale of $\mu=$1~GeV,
where the central values of Gegenbauer moments given in
Eqs.(\ref{eq:gvalues1}) and (\ref{eq:gvalues2}) are used. $u$
($\bar u\equiv 1-u$) is the meson momentum fraction carried by the
quark (antiquark). The solid, dashed and dot-dashed curves
correspond to $b_1(1235), h_1(1380)$ and $\overline K_{1B}$,
respectively.}}}
\end{center}
\end{figure}

We have calculated the first few Gegenbauer moments of leading
twist light-cone distribution amplitudes of $1^1P_1$ mesons using
the QCD sum rule technique. The models for light-cone distribution
amplitudes depend on the Gegenbauer moments of the truncated
conformal expansion. Taking into account $a_{0,1,2}^{\parallel,A}$
and $a_{1,2,3}^{\perp, A}$, we show the light-cone distribution
amplitudes in Fig.~\ref{fig:lcdas}. $\Phi_\parallel^A(u)$ is
asymmetric under $u \leftrightarrow 1-u$ if neglecting SU(3)
breaking effects. In particular, we obtain sizable magnitudes for
the first Gegenbauer moments of $\Phi_\parallel^A(u)$: $f_A
a_1^{\parallel,A}(1~{\rm GeV})\simeq (-0.23\sim -0.39)$~GeV, which
could greatly enhance the longitudinal branching ratios of
factorization-suppressed $B \to h_1(1380)  K^{(*)}, b_1(1235)
K^{(*)}$ modes~\cite{Yang:2005tv}. Unfortunately, it seems to be
impossible to obtain reliable estimates for $f_A
a_l^{\parallel,A}$ with $l\geq 3$.

Recently, Belle has measured $B^- \to K_1^- (1270) \gamma$ and
given an upper bound on $B^- \to K_1^-(1400)
\gamma$~\cite{Yang:2004as}. Interestingly, the recent
calculations~\cite{Safir:2001cd,Kwon:2004ri} of adopting LCSR
(light-cone sum rule) form factors gave too small predictions for
${\cal B}(B^- \to K_1^- (1270) \gamma)$ as compared with the data.
Since the physical states $K_1(1270)$ and $K_1(1400)$ are the
mixture of $K_{1A}$ and $K_{1B}$ which are respectively the pure
$1^3P_1$ and $1^1P_1$ states, the light-cone distribution
amplitudes of $K_{1A}$ and $K_{1B}$ are relevant to the results of
$B\to K_1(1270)$ and $K_1(1400)$ transition form factors. It is
known that for $K_{1B}$, $\Phi_\parallel$ is antisymmetric, while
$\Phi_\perp$ is symmetric in the SU(3) limit due to G-parity.
Nevertheless, for $K_{1A}$, $\Phi_\parallel$ becomes symmetric,
while $\Phi_\perp$ is antisymmetric. The above properties were not
studied in the literature. These related researches will be
published in the near future~\cite{kcymoments}.

\subsubsection*{Acknowledgments}

 This work was supported in
part by the National Science Council of R.O.C. under Grant Nos:
NSC93-2112-M-033-004 and NSC94-2112-M-033-001.

 \end{document}